\newcommand{\roughly}[1]%
    {{\mathrel{\raise.3ex\hbox{$#1$\kern-.75em\lower1ex\hbox{$\sim$}}}}}
\newcommand{\lsim}{\mathrel{\roughly<}}
\newcommand{\gsim}{\mathrel{\roughly>}}

\documentclass[12pt]{article}
\usepackage{paper2e}
\usepackage{epsfig}
\usepackage{latexsym}

\begin{document}
\begin{titlepage}

\preprint{UTAP-569, RESCEU-26/06}

\title{Towards a Higgs phase of gravity in string theory}

\author{Shinji Mukohyama}

\address{Department of Physics and 
Research Center for the Early Universe,\\
The University of Tokyo, Tokyo 113-0033, Japan}

\begin{abstract}
 We consider a braneworld scenario with a black brane parallel to our 
 brane, aiming towards a Higgs phase of gravity in string theory. The
 existence of the black brane spontaneously breaks the Lorentz symmetry
 on our brane but preserves the rotational and translational
 invariance. If all moduli are stabilized then this should lead to a
 Higgs phase of gravity. We investigate moduli stabilization by using a
 KKLT-type moduli potential in the context of type IIB warped flux
 compactification. 
\end{abstract}

\end{titlepage}

\section{Introduction}

Dark energy and dark matter are two major mysteries in modern
cosmology. Although more than $90\%$ of our universe is filled with
them, we do not know what they really are. In this situation it seems
rather natural to ask whether we can modify gravity in the infrared (IR)
to address those mysteries.

However, modifying gravity in the IR is not easy. For example, in
massive gravity~\cite{Fierz:1939ix} and the DGP brane
model~\cite{Dvali:2000hr}, it is known that a scalar degree of freedom
becomes strongly coupled at a rather low energy
scale~\cite{Arkani-Hamed:2002sp,Luty:2003vm}. If gravity is to be
modified at and beyond the present Hubble distance then the scalar
degree of freedom becomes strongly coupled and its dynamics is dominated 
by quantum corrections at and within $\sim 1000km$. This is similar to
what happens to a massive gauge theory, but the situation is worse. In a
massive gauge theory one can decouple the strongly coupled sector from
the rest of the world by tuning the gauge coupling to a small value. On
the other hand, in the massive gravity and the DGP model, the strength
of interactions between the strongly coupled sector and the rest of the
world is set by the Newton's constant and cannot be fine-tuned
away. This means that, as soon as the strong coupling appears,
gravitational phenomena cannot be properly described by those theories
without knowledge about yet-unknown ultraviolet (UV) completion. Since
this issue in the UV is originated from the properties in the IR, it is
not totally clear whether it is possible to UV complete those theories
without affecting the IR behaviors.

In a gauge theory, it is the Higgs mechanism that gives a mass term to a
gauge boson and modifies a force law in a theoretically controllable
way. Thus, it is natural to apply the idea of the Higgs mechanism to
gravity to modify its IR behavior. Since the symmetry relevant to
gravity is diffeomorphism invariance, the Higgs mechanism for gravity
should spontaneously break at least part of the diffeomorphism
invariance. Note that in gravitational theories, Lorentz breaking
inevitably implies breaking diffeomorphism invariance and, thus, a Higgs 
phase of gravity.

The simplest Higgs phase of gravity is the ghost
condensation~\cite{Arkani-Hamed:2003uy}, including just one
Nambu-Goldstone (NG) boson. Because of the simplicity and the
universality of the low energy effective field theory (EFT), it is
worthwhile investigating properties of the ghost condensation, whether
or not it leads to interesting physical phenomena. Fortunately, 
it turns out that the physics in the simplest Higgs phase of gravity is
extremely rich and interesting. They include IR modification of 
gravity~\cite{Arkani-Hamed:2003uy}, a new spin-dependent 
force~\cite{Arkani-Hamed:2004ar},  a qualitatively different picture of 
inflationary de Sitter phase~\cite{Arkani-Hamed:2003uz,Senatore:2004rj}, 
effects of moving sources~\cite{Dubovsky:2004qe,Peloso:2004ut}, 
intriguing nonlinear dynamics~\cite{Krotov:2004if,Arkani-Hamed:2005gu},
properties of black 
holes~\cite{Frolov:2004vm,Mukohyama:2005rw,Dubovsky:2006vk},
implications to galaxy rotation
curves~\cite{Kiselev:2004bh,Kiselev:2005sv,Kiselev:2006gc}, dark energy
models~\cite{Piazza:2004df,Krause:2004bu,deRham:2006pe}, stable
violation of null energy
condition~\cite{Creminelli:2006xe}, cosmological
perturbations~\cite{Mukohyama:2006be}, other classical 
dynamics~\cite{Anisimov:2004sp,Mann:2005jz}, attempts towards UV 
completion~\cite{Graesser:2005ar,O'Connell:2006de}, and so on. 
In the ghost condensation the strength of interactions between the NG
boson and the rest of the world is controlled by not only the Newton's
constant but also the scale of spontaneous Lorentz breaking ${\cal M}$
in such a way that the interactions are completely turned off in the
limit ${\cal M}/M_{Pl}\to 0$. Therefore, general relativity is safely
recovered in this limit. In ref.~\cite{Arkani-Hamed:2005gu} it was
argued that the theory is compatible with all current experimental
observations if ${\cal M}$ is lower than 
$\sim 100\ensuremath{\mathrm{~GeV}}$. Therefore, the ghost condensation 
opens up a number of new avenues for attacking cosmological problems,
including inflation, dark matter and dark energy.

While phenomenologies of the ghost condensation are still under
investigation, it is certainly important to seek a possible UV
completion.

To realize the ghost condensation without fine-tuning, we need to
spontaneously break the $4$-dimensional diffeomorphism invariance times
a global shift symmetry down to the $3$-dimensional spatial
diffeomorphism invariance times an unbroken global shift symmetry, where
the latter global shift is a combination of the former global shift and
the time shift. However, it is generally believed that all symmetries in
string theory are gauged. Therefore, it seems more plausible to obtain
the ghost condensation as the neutral limit of the gauged version of the
ghost condensation, i.e. the gauged ghost
condensation~\cite{Cheng:2006us}. To obtain the gauged ghost
condensation from the ghost condensation we replace the global shift
symmetry with a minimal gauge symmetry, i.e. $U(1)$ gauge symmetry, so
that no global symmetry is needed. The ghost condensation can be
obtained from the gauged ghost condensation if we can fine-tune the
gauge coupling to a sufficiently small value. Some of low energy
properties of the gauged ghost condensation, including the bound 
${\cal M}\lsim \mathrm{Min}(10^{12}\mathrm{~GeV},\, 
g^2\,10^{15}\mathrm{~GeV})$ 
for $g^2>g_c^2$, have been investigated in \cite{Cheng:2006us}. Here,
$g$ is the gauge coupling and $g_c={\cal M}^2/2M_{Pl}^2$.

The purpose of this paper is, as a step towards a Higgs phase of gravity
in string theory, to consider a braneworld scenario in which a black
brane parallel to our brane breaks the Lorentz symmetry on our brane. If
all moduli are stabilized then this leads to a Higgs phase of gravity in
our world. From symmetry arguments, we expect that this is closely
related to the gauged ghost condensation.

The rest of this paper is organized as follows. In Sec.~\ref{sec:GGC} we
review the gauged ghost condensation by deriving its low energy
EFT. This makes it clear that the structure of the EFT is determined
solely by the symmetry breaking pattern. In other words, different
setups should result in low energy EFTs with the same structure if the
symmetry breaking pattern is the same. In
Sec.~\ref{Jeans-GL-instability} we investigate the dispersion relation
of the NG boson coupled with gravity and show that it exhibits
Jeans-like instability if the gauge coupling in the EFT is smaller than
a critical value. We also point out a number of similarities between the
Jeans-like instability of the NG boson and the Gregory-Laflamme (GL)
instability of black branes~\cite{Gregory:1993vy,Gregory:1994bj}. Those
similarities motivate a braneworld setup considered in 
Sec.~\ref{sec:string}. In Sec.~\ref{sec:black-brane} it is argued that, 
in (higher-dimensional) general relativity, a brane with codimension
more than $2$ should form a black brane if the brane is sufficiently
thin. The string theory version of this statement is the correspondence
principle for $D$-branes and black branes~\cite{Horowitz:1996nw}, which
plays a central role in the braneworld setup. In
Sec.~\ref{sec:black-brane} we also see close connections between
extremality of a black brane and the Lorentz symmetry along the
world-volume of the black brane. Motivated by various considerations in
the preceding sections, a braneworld setup in string theory is
considered in Sec.~\ref{sec:string}. Sec.~\ref{sec:summary} is devoted
to a summary of this paper and discussions.

\section{Gauged ghost condensation}
\label{sec:GGC}

The ghost condensation is the simplest Higgs phase of gravity in the
sense that it contains only one Nambu-Goldstone boson. The gauged ghost 
condensation is obtained by gauging the global shift symmetry of the
ghost condensation~\cite{Cheng:2006us}. In this section we give an
alternative derivation of the effective field theory (EFT) of the gauged
ghost condensation based on a symmetry breaking pattern.

For the gauged ghost condensation, we assume the following symmetry
breaking pattern. 
\begin{itemize}
 \item[(i)] The $4$-dimensional diffeomorphism invariance and a
	    $U(1)$ gauge symmetry is spontaneously broken down. The
	    residual symmetries are the $3$-dimensional spatial
	    diffeomorphism invariance, the time reversal symmetry and a
	    $U(1)$ gauge symmetry. The residual $U(1)$ is a combination
	    of the broken $U(1)$ and the broken time
	    reparameterization. 
\item[(ii)] The background spacetime metric is maximally symmetric,
	    either Minkowski or de Sitter. 
\end{itemize}
Our strategy here is to write down the most general action invariant
under the residual symmetries. This gives the effective action in the
unitary gauge. After that, the action in a general gauge is obtained by
gauge transformation. The gauge parameter associated with the broken
symmetry actually appears in the action and is identified as the NG
boson associated with the spontaneous symmetry breaking.

For simplicity let us consider a Minkowski background plus perturbation:
$g_{\mu\nu}=\eta_{\mu\nu}+h_{\mu\nu}$. We also consider a $U(1)$ gauge
field $a_{\mu}$. The general infinitesimal diffeomorphism and $U(1)$
gauge transformations are 
$\delta h_{\mu\nu}=\partial_{\mu}\xi_{\nu}+\partial_{\nu}\xi_{\mu}$ and 
$\delta a_{\mu}=\partial_{\mu}\chi$, where the $4$-vector $\xi^{\mu}$
and the scalar $\chi$ represent gauge degrees of freedom. The residual
$3$-dimensional spatial diffeomorphism is given by 
%
\begin{equation}
 \xi^0=0, \quad \xi^i = \xi^i(t,\vec{x}), \quad \chi = 0,
\end{equation}
and transforms the metric perturbation and the $U(1)$ gauge field as
%
\begin{equation}
 \delta h_{00}=0, \quad \delta h_{0i}=\partial_0\xi_i, \quad
  \delta h_{ij}=\vec{\nabla}_i\xi_j+\vec{\nabla}_j\xi_i, \quad
  \delta a_0 = \delta a_i = 0. 
  \label{eqn:3d-diffeo}
\end{equation}
The residual $U(1)$ gauge symmetry is characterized by 
%
\begin{equation}
 \xi^0\, (=-\xi_0) =-\frac{g}{{\cal M}}\chi(t,\vec{x}), \quad \xi^i = 0,
  \quad \chi = \chi(t,\vec{x}), 
\end{equation}
and transforms the fields as
%
\begin{equation}
 \delta h_{00} = \frac{2g}{{\cal M}}\partial_0\chi, \quad
  \delta h_{0i} = \frac{g}{{\cal M}}\vec{\nabla}_i\chi, \quad
  \delta h_{ij} = 0, \quad
  \delta a_0 = \partial_0\chi, \quad
  \delta a_i = \vec{\nabla}_i\chi,
  \label{eqn:residual-U1}
\end{equation}
where ${\cal M}$ is the scale of the spontaneous Lorentz breaking and 
$g$ is the gauge coupling. On the other hand, the choice
%
\begin{equation}
 \xi^0\, (=-\xi_0)=\pi(t,\vec{x}), \quad \xi^i = 0, \quad  \chi = 0 
\end{equation}
corresponds to the broken symmetry and the fields are transformed as
%
\begin{equation}
 \delta h_{00} = -2\partial_0\pi, \quad
  \delta h_{0i} = -\vec{\nabla}_i\pi, \quad
  \delta h_{ij} = 0, \quad
  \delta a_0 = \delta a_i = 0.
  \label{eqn:broken-sym}
\end{equation}

Now let us seek terms invariant under the both residual gauge
transformations (\ref{eqn:3d-diffeo}) and (\ref{eqn:residual-U1}). They
must begin at quadratic order since we assumed that the Minkowski 
background is a solution to the equation of motion. The leading term
(without derivatives acted on the metric perturbations) is 
$\int d\vec{x}^3dt\, \tilde{h}_{00}^2$, where 
%
\begin{equation}
 \tilde{h}_{00} \equiv h_{00} - \frac{2g}{{\cal M}}a_0.
\end{equation}
This is indeed invariant under both residual gauge transformations
(\ref{eqn:3d-diffeo}) and (\ref{eqn:residual-U1}). From this term, we
can obtain the corresponding term in the effective action for the NG
boson $\pi$. Since $\tilde{h}_{00}\to \tilde{h}_{00}-2\partial_0\pi$
under the broken symmetry transformation (\ref{eqn:broken-sym}), by
promoting $\pi$ to a physical degree of freedom we obtain the term 
$\int d\vec{x}^3dt\, (\tilde{h}_{00}-2\partial_0\pi)^2$. This includes
a time kinetic term for $\pi$ as well as mixing terms. At this point we  
wonder if we can get the usual space kinetic term $(\vec{\nabla}\pi)^2$
or not. The only possibility would be from $(h_{0i})^2$ since 
$h_{0i}\to h_{0i}-\vec{\nabla}_i\pi$ under the broken symmetry
transformation (\ref{eqn:broken-sym}). However, this term is not
invariant under either (\ref{eqn:3d-diffeo}) or (\ref{eqn:residual-U1}),
and, thus, cannot enter the effective action. Actually, by acting
derivatives on the metric components, we can find combinations
manifestly invariant under the spatial diffeomorphism
(\ref{eqn:3d-diffeo}). They are made of the geometrical quantity called 
extrinsic curvature. The extrinsic curvature $K_{ij}$ of a constant time
surface is 
$K_{ij}=(\partial_0h_{ij}-\vec{\nabla}_ih_{0j}-\vec{\nabla}_jh_{0j})/2$ 
in the linear order and transforms as a tensor under the spatial
diffeomorphism. Indeed, it is invariant under the spatial diffeomorphism
(\ref{eqn:3d-diffeo}) since the background value of the extrinsic
curvature vanishes. Although it is not invariant under the residual
$U(1)$ gauge transformation (\ref{eqn:residual-U1}), it is possible to 
compensate the transformation of $K_{ij}$ with that of derivatives of
the $U(1)$ gauge field. Indeed, the combination 
%
\begin{equation}
 \tilde{K}_{ij} \equiv K_{ij} + \frac{g}{{\cal M}}\vec{\nabla}_{(i}a_{j)}
\end{equation}
is invariant under both (\ref{eqn:3d-diffeo}) and
(\ref{eqn:residual-U1}), where 
$\vec{\nabla}_{(i}a_{j)}\equiv(\vec{\nabla}_ia_j+\vec{\nabla}_ja_i)/2$.
Thus, $\int d\vec{x}^3dt\, \tilde{K}^2$ and 
$\int d\vec{x}^3dt\, \tilde{K}^{ij}\tilde{K}_{ij}$ can be used in
the action. Since 
$\tilde{K}_{ij}\to \tilde{K}_{ij}+\vec{\nabla}_i\vec{\nabla}_j\pi$ under
the broken symmetry (\ref{eqn:broken-sym}), we obtain 
$\int d\vec{x}^3dt\, (\tilde{K}+\vec{\nabla}^2\pi)^2$ and 
$\int d\vec{x}^3dt\, (\tilde{K}^{ij}+\vec{\nabla}^i\vec{\nabla}^j\pi)
(\tilde{K}_{ij}+\vec{\nabla}_i\vec{\nabla}_j\pi)$. 
The field strength
$F_{\mu\nu}=\partial_{\mu}a_{\nu}-\partial_{\nu}a_{\mu}$ is also
invariant under both (\ref{eqn:3d-diffeo}) and
(\ref{eqn:residual-U1}). Thus, $\int d\vec{x}^3dt\, F_0^{\ i}F_{0i}$ and 
$\int d\vec{x}^3dt\, F^{ij}F_{ij}$ can enter the action. These do not generate
terms involving the NG boson. Note that terms like
$\int d\vec{x}^3dt\, \tilde{K}\tilde{h}_{00}$ are forbidden by the time
reversal symmetry.

Combining these terms, we obtain the effective action 
$\int d\vec{x}^3dt\,L$, where
%
\begin{eqnarray}
 L & = &
  \frac{{\cal M}^4}{2}
  \left(\partial_0\pi+\frac{g}{{\cal M}}a_0-\frac{1}{2}h_{00}\right)^2 
  - \frac{\alpha_1{\cal M}^2}{2}
  \left(\vec{\nabla}^2\pi+\frac{g}{{\cal M}}\vec{\nabla}^ia_i+K\right)^2
  \nonumber\\
 & &
  - \frac{\alpha_2{\cal M}^2}{2}
  \left(\vec{\nabla}^i\vec{\nabla}^j\pi
   +\frac{g}{{\cal M}}\vec{\nabla}^{(i}a^{j)}
   + K^{ij}\right)
  \left(\vec{\nabla}_i\vec{\nabla}_j\pi
   +\frac{g}{{\cal M}}\vec{\nabla}_{(i}a_{j)}
   + K_{ij}\right)
  \nonumber\\
 & &
  + \frac{1}{2}F_0^{\ i}F_{0i}
  - \frac{\gamma_1}{4}F^{ij}F_{ij} + \cdots.
  \label{eqn:effective-action-gcc}
\end{eqnarray}
Here, $\alpha_{1,2}$ and $\gamma_1$ are dimensionless constants of order 
unity and we have normalized $\pi$, $a_{\mu}$ and $g$ so that 
the coefficients of the first and the fourth terms become ${\cal M}^4/2$ 
and $1/2$, respectively.

In deriving the effective action (\ref{eqn:effective-action-gcc}), all we
needed was the symmetry breaking pattern. As intended, this action
agrees with the effective action obtained in ref.~\cite{Cheng:2006us} by
simply gauging the shift symmetry in the ghost condensation.

This action has the following symmetry:
%
\begin{equation}
 h_{\mu\nu}\to h_{\mu\nu}+ \partial_{\mu}\xi_{\nu}+\partial_{\nu}\xi_{\mu},
  \quad
  a_{\mu}\to a_{\mu}+\partial_{\mu}\chi, \quad
  \pi \to \pi + \xi_0 - \frac{g}{{\cal M}}\chi,
\end{equation}
where the gauge parameters $\xi^{\mu}$ and $\chi$ depend on $t$ and
$\vec{x}$. By using $\chi$, we can set $a_0=0$ and obtain the effective
action
%
\begin{eqnarray}
 L & = &
  \frac{{\cal M}^4}{2}
  \left(\partial_0\pi-\frac{1}{2}h_{00}\right)^2 
  - \frac{\alpha_1{\cal M}^2}{2}
  \left(\vec{\nabla}^2\pi+\frac{g}{{\cal M}}\vec{\nabla}^ia_i+K\right)^2
  \nonumber\\
 & &
  - \frac{\alpha_2{\cal M}^2}{2}
  \left(\vec{\nabla}^i\vec{\nabla}^j\pi
   +\frac{g}{{\cal M}}\vec{\nabla}^{(i}a^{j)}
   + K^{ij}\right)
  \left(\vec{\nabla}_i\vec{\nabla}_j\pi
   +\frac{g}{{\cal M}}\vec{\nabla}_{(i}a_{j)}
   + K_{ij}\right)
  \nonumber\\
 & &
  + \frac{1}{2}\partial_0a^i\partial_0a_i
  - \frac{\gamma_1}{4}F^{ij}F_{ij} + \cdots.
  \label{eqn:effective-action}
\end{eqnarray}
This action has the following symmetry:
%
\begin{equation}
 h_{\mu\nu}\to h_{\mu\nu}+ \partial_{\mu}\xi_{\nu}+\partial_{\nu}\xi_{\mu},
  \quad
  a_i\to a_i+\vec{\nabla}_i\chi^{(3)}, \quad
  \pi \to \pi + \xi_0 - \frac{g}{{\cal M}}\chi^{(3)},
  \label{eqn:effective-action-symmetry}
\end{equation}
where $\chi^{(3)}$ depends on $\vec{x}$ but is independent of $t$.

If the gauge coupling in the EFT is small enough then this reduces to
the ghost condensation~\cite{Arkani-Hamed:2003uy}: 
%
\begin{eqnarray}
 L_{g\to 0} & = &
  \frac{{\cal M}^4}{2}
  \left(\partial_0\pi-\frac{1}{2}h_{00}\right)^2 
  - \frac{\alpha_1{\cal M}^2}{2}
  \left(\vec{\nabla}^2\pi+K\right)^2
  \nonumber\\
 & &
  - \frac{\alpha_2{\cal M}^2}{2}
  \left(\vec{\nabla}^i\vec{\nabla}^j\pi
   + K^{ij}\right)
  \left(\vec{\nabla}_i\vec{\nabla}_j\pi
   + K_{ij}\right)
  + \cdots,
  \label{eqn:effective-action-g=0}
\end{eqnarray}
This action includes leading terms of the quadratic order
only. Actually, it is possible to show that nonlinear terms are
irrelevant at low energies. To begin with, suppose that the energy is
scaled as $E\mapsto sE$. Then, the time interval $dt$ scales as
$dt\mapsto s^{-1}dt$. To determine the scaling rule for the spatial
interval $d\vec{x}$ and the NG boson $\pi$, we demand that the leading 
action $\int d\vec{x}^3dt\,L_{g\to 0}$ with $h_{\mu\nu}=0$ be 
invariant under the scaling. The results are summarized as  
%
\begin{eqnarray}
 E & \mapsto & sE, \nonumber\\
 dt & \mapsto & s^{-1}dt, \nonumber\\
 d\vec{x} & \mapsto & s^{-1/2}d\vec{x}, \nonumber\\
 \pi & \mapsto & s^{1/4}\pi.
\end{eqnarray}
Note that the scaling dimension of $\pi$ is not equal to the mass
dimension $1$ but is $1/4$~\cite{Arkani-Hamed:2003uy}. By using this
scaling, it is easy to identify the scaling dimensions of any nonlinear
operators. The leading nonlinear operator 
%
\begin{equation}
 \int d\vec{x}^3dt\, {\cal M}^4 \dot{\pi}(\vec{\nabla}\pi)^2
\end{equation}
scales as $s^{1/4}$ and is irrelevant at low energies. All other
nonlinear operators have scaling dimension $1/4$ or higher and, thus,
are irrelevant. Therefore, those nonlinear operators become less and
less important as energies and momenta become low compared with the
scale ${\cal M}$. Low energy dynamics of the NG boson is, thus, well
described by the effective action.

Going back to the action (\ref{eqn:effective-action}) for a
non-vanishing $g$, the above power-counting analysis shows that at low
energies and momenta, all nonlinear operators made of $\pi$ and its
derivatives are irrelevant compared with the first three terms. On the
other hand, the fourth and fifth terms in (\ref{eqn:effective-action}) 
give leading kinetic terms for $a_i$. Therefore, the action
(\ref{eqn:effective-action}) well describes the low energy dynamics of
the system.

In summary, (\ref{eqn:effective-action}) is the low energy effective of
the gauged ghost condensation characterized by the symmetry breaking
pattern in the beginning of this section. The structure of the effective
action is universal since it is determined solely by the symmetry
breaking pattern. Different setups of Higgs phase of gravity should
result in low energy EFTs of the same structure if their symmetry
breaking patters are the same. If the gauge coupling $g$ is small enough
then the effective action is reduced to (\ref{eqn:effective-action-g=0})
and agrees with that of the ghost condensation.

\section{Jeans-like instability and Gregory-Laflamme instability} 
\label{Jeans-GL-instability}

The EFT (\ref{eqn:effective-action}) should be coupled to Einstein
gravity. As we shall see below, qualitative behavior of the coupled
system depends on whether the gauge coupling $g$ is larger than a
critical value $g_c$ or smaller~\cite{Cheng:2006us}. For $g^2>g_c^2$,
the linear perturbation around Minkowski background is stable. On the
other hand, for $g^2<g_c^2$, the coupled system exhibits a Jeans-like
instability for long wavelength modes as in the ghost
condensation~\cite{Arkani-Hamed:2003uy}. In the end of this section we
shall point out similarities between this Jeans-like instability and the
Gregory-Laflamme (GL) instability~\cite{Gregory:1993vy,Gregory:1994bj}
of black branes. This similarity is what will motivate us to consider a
scenario for a Higgs phase of gravity in string theory explained in
Sec.~\ref{sec:string}.

Let us now derive the dispersion relation for the NG boson described by
(\ref{eqn:effective-action}). Since the $3$-dimensional spatial
diffeomorphism invariance is preserved, linear perturbations around the
background can be decomposed into scalar, vector and tensor type,
according to their transformation properties under the spatial
diffeomorphism. In this section we consider scalar-type perturbation.

For scalar-type perturbation, $a_i$ is written as
$a_i=\vec{\nabla}_ia_L$, where $a_L$ depends on $t$ and $\vec{x}$. By 
setting $\alpha_2=0$ for simplicity, the action
(\ref{eqn:effective-action}) then reduces to 
%
\begin{equation}
 L = \frac{{\cal M}^4}{2}
  \left(\partial_0\pi-\frac{1}{2}h_{00}\right)^2 
  - \frac{\alpha_1{\cal M}^2}{2}
  \left(\vec{\nabla}^2\pi+\frac{g}{{\cal M}}\vec{\nabla}^2a_L+K\right)^2
  + \frac{1}{2}(\partial_0\vec{\nabla}a_L)^2.
\end{equation}
Thus, the equations of motion derived from the total action 
$\int d\vec{x}^3dt\, (L_{EH}+L)$ are
%
\begin{eqnarray}
 M_{\rm Pl}^2 G_{00} 
  + {\cal M}^4\left(\partial_0\pi-\frac{1}{2}h_{00}\right) & = & 0,
  \nonumber\\
 M_{\rm Pl}^2 G_{0i} - \alpha_1{\cal M}^2\vec{\nabla}_i
    \left(\vec{\nabla}^2\pi+\frac{g}{{\cal M}}\vec{\nabla}^2a_L+K\right)
    & = & 0,  \nonumber\\
 M_{\rm Pl}^2 G_{ij} - \alpha_1{\cal M}^2\partial_0
  \left(\vec{\nabla}^2\pi+\frac{g}{{\cal M}}\vec{\nabla}^2a_L+K\right)
  \delta_{ij} & = & 0, \nonumber\\
 \partial_0^2a_L - \alpha_1 g{\cal M}
  \left(\vec{\nabla}^2\pi+\frac{g}{{\cal M}}\vec{\nabla}^2a_L+K\right)
  & = & 0, \nonumber\\
 \partial_0\left(\partial_0\pi-\frac{1}{2}h_{00}\right)
  + \frac{\alpha_1}{{\cal M}^2}\vec{\nabla}^2
  \left(\vec{\nabla}^2\pi+\frac{g}{{\cal M}}\vec{\nabla}^2a_L+K\right)
  & = & 0.
\end{eqnarray}
In the longitudinal gauge
%
\begin{equation}
 ds_4^2 = -(1+2\Phi)dt^2 + (1-2\Psi)d\vec{x}^2,
\end{equation}
the Einstein tensor and the extrinsic curvature are
%
\begin{eqnarray}
 G_{00} & = & 2\vec{\nabla}^2\Psi, \nonumber\\
 G_{0i} & = & 2\partial_0\vec{\nabla}_i\Psi, \nonumber\\
 G_{ij} & = & 2\left[\partial_0^2\Psi +
		\frac{1}{3}\vec{\nabla}^2(\Phi-\Psi)\right]\delta_{ij}
 - \left(\vec{\nabla}_i\vec{\nabla}_j
    -\frac{1}{3}\vec{\nabla}^2\delta_{ij}\right)(\Phi-\Psi), \nonumber\\
  K_{ij} & = &-\partial_0\Psi\delta_{ij}.
\end{eqnarray}
Thus, the traceless part of the third equation of motion implies
$\Phi=\Psi$. The rest of the equations of motion become
%
\begin{eqnarray}
 2M_{\rm Pl}^2 \vec{\nabla}^2\Phi
  + {\cal M}^4(\partial_0\pi+\Phi) & = & 0,
  \nonumber\\
 2M_{\rm Pl}^2\partial_0\vec{\nabla}_i\Phi
  - \alpha_1{\cal M}^2\vec{\nabla}_i
    \left(\vec{\nabla}^2\pi+\frac{g}{{\cal M}}\vec{\nabla}^2a_L
     -3\partial_0\Phi\right)
    & = & 0,  \nonumber\\
 2M_{\rm Pl}^2 \partial_0^2\Phi - \alpha_1{\cal M}^2\partial_0
  \left(\vec{\nabla}^2\pi+\frac{g}{{\cal M}}\vec{\nabla}^2a_L
   -3\partial_0\Phi\right) & = & 0, \nonumber\\
 \partial_0^2a_L - \alpha_1 g{\cal M}
  \left(\vec{\nabla}^2\pi+\frac{g}{{\cal M}}\vec{\nabla}^2a_L
   -3\partial_0\Phi\right) & = & 0, \nonumber\\
 \partial_0(\partial_0\pi+\Phi)
  + \frac{\alpha_1}{{\cal M}^2}\vec{\nabla}^2
  \left(\vec{\nabla}^2\pi+\frac{g}{{\cal M}}\vec{\nabla}^2a_L
   -3\partial_0\Phi\right) & = & 0. 
\end{eqnarray}
The second and third equations imply that
%
\begin{equation}
 \partial_0\Phi = \frac{\alpha_1g_c^2}{1+3\alpha_1g_c^2}
    \left(\vec{\nabla}^2\pi+\frac{g}{{\cal M}}\vec{\nabla}^2a_L\right),
    \label{eqn:d0Phi}
\end{equation}
where
%
\begin{equation}
 g_c^2 = \frac{{\cal M}^2}{2M_{\rm Pl}^2}.
  \label{eqn:def-gc}
\end{equation}
The physical meaning of $g_c$ will be clarified soon. By substituting 
this to the fifth equation, we obtain 
%
\begin{equation}
 \partial_0^2\pi +
  \frac{\alpha_1}{1+3\alpha_1g_c^2}
  \left(g_c^2+\frac{\vec{\nabla}^2}{{\cal M}^2}\right)
  \left(\vec{\nabla}^2\pi+\frac{g}{{\cal M}}\vec{\nabla}^2a_L\right)
  = 0.
\end{equation}
Finally, by acting the operator 
$[\partial_0^2-\alpha_1g^2\vec{\nabla}^2/(1+3\alpha_1g_c^2)]$ on this
equation and using the fourth equation (with the substitution of
(\ref{eqn:d0Phi})), we obtain 
%
\begin{equation}
 \partial_0^2
  \left[\partial_0^2
   + \frac{\alpha_1}{1+3\alpha_1g_c^2}
   \left(-g^2+g_c^2+\frac{\vec{\nabla}^2}{{\cal M}^2}\right)
   \vec{\nabla}^2\right]\pi = 0.
\end{equation}
Thus, by neglecting $3\alpha_1g_c^2$ ($\ll 1$) in the denominator, this
equation implies the following dispersion relation: 
%
\begin{equation}
 \omega^2 = \alpha_1(g^2-g_c^2)\vec{k}^2 
  + \frac{\alpha_1}{{\cal M}^2}\vec{k}^4,
  \label{eqn:dispersion-relation}
\end{equation}
where $g_c$ is defined by (\ref{eqn:def-gc}). For the stability of modes
with high momenta, we assume that $\alpha_1>0$. 
When $g^2\ll g_c^2$, the gauged ghost condensation reduces to the ghost
condensation and the dispersion relation becomes
%
\begin{equation}
 \omega^2 = -\frac{\alpha_1{\cal M}^2}{2M_{\rm Pl}^2}\vec{k}^2 
  + \frac{\alpha_1}{{\cal M}^2}\vec{k}^4 \quad \mbox{for }g^2\ll g_c^2.
  \label{eqn:dispersion-relation-gc}
\end{equation}

From the dispersion relation (\ref{eqn:dispersion-relation}) it is easy
to see that the NG boson coupled with the $U(1)$ field and gravity is
stable if $g^2>g_c^2$. On the other hand, for $g^2<g_c^2$, modes with
long wavelength are unstable. For a usual fluid with the background
energy density $\rho_0$, the dispersion relation is 
$\omega^2=c_s^2{\bf k}^2-\omega_J^2$, where $c_s$ is the sound speed and 
$\omega_J^2=4\pi G_N\rho_0$, and long-scale modes with 
$c_s^2{\bf k}^2<\omega_J^2$ have instability called Jeans instability. 
The Jeans instability is originated from the attractive nature of
gravity, and is an important instability rather than catastrophe since
it contributes to the structure formation in the universe. The
long-scale instability indicated by the dispersion relation
(\ref{eqn:dispersion-relation}) for $g^2<g_c^2$ is an analog of the
Jeans instability. In the limit $g^2\ll g_c^2$, from
(\ref{eqn:dispersion-relation-gc}) one can see that the length and time
scales characterizing the most unstable mode are 
%
\begin{equation}
 r_c\simeq \frac{M_{\rm Pl}}{{\cal M}^2}, \quad
  t_c\simeq \frac{M_{\rm Pl}^2}{{\cal M}^3} \quad 
  \mbox{for } g^2\ll g_c^2,
\end{equation}
and are much longer than $1/{\cal M}$. For larger $g^2$ ($<g_c^2$), the
length and time scales are even longer.

Thus, the physical meaning of $g_c$ is the critical value of the gauge
coupling below which the NG boson exhibits the Jeans-like
instability. For small gauge coupling $g^2<g_c^2$, the attractive nature
of gravity dominates over the repulsive $U(1)$ gauge force. This is the
reason for the Jeans-like instability, which reflects the attractive
nature of gravity. On the other hand, for large gauge coupling 
$g^2>g_c^2$, the repulsive nature of the $U(1)$ gauge force dominates
over the attractive nature of gravity and the Jeans-like instability
disappears. Of course, even for $g^2<g_c^2$, the Jeans-like instability
disappears in the expanding universe if the Hubble expansion rate is
large enough.

Gregory-Laflamme (GL) instability~\cite{Gregory:1993vy,Gregory:1994bj}
of black brane solutions is a classical instability in which the mass of
the black brane tends to clump non-uniformly. Intriguingly, as we shall
point out in the following, there are a couple of similarities to the
Jeans-like instability of the NG boson explained above.

First, both instabilities are for long wavelength modes. Modes with
wavelength shorter than a critical length are stable in both
cases. Indeed, the dispersion relation for the GL instability is
qualitatively the same as (\ref{eqn:dispersion-relation}) with
$g^2<g_c^2$ as one can see from Fig.1 of \cite{Gregory:1993vy} and Fig.6 
of \cite{Gregory:1994bj}. Note that the momentum $\mu$ along the
world-volume of the black brane must be replaced by $\sqrt{\vec{k}^2}$
and the growth rate $\Omega$ must be replaced by $\sqrt{-\omega^2}$ in
order to compare these two dispersion relations.

Second, both instabilities disappear when the Lorentz symmetry is
recovered. The recovery of the Lorentz symmetry corresponds to the limit 
${\cal M}/M_{\rm Pl}\to 0$ in the above EFT coupled to gravity, and in
this limit the Jeans-like instability indeed disappears ($r_c\to\infty$
and $t_c\to\infty$) even for $g^2\ll g_c^2$. On the other hand, as we
shall see in Sec.~\ref{sec:black-brane} and
Appendix~\ref{app:four-para-sol}, a non-extremal black brane tends to
break the Lorentz symmetry along its world-volume: a non-extremal black
brane has a preferred frame in which it is at rest, and a Lorentz boost
even in the direction parallel to its world-volume does not preserve the
form of the metric. However, as we shall see explicitly for a
four-parameter family of black $p$-brane solutions, when the Lorentz
symmetry is recovered, a black brane becomes extremal and
BPS-saturated. It is thought that extremal black $p$-branes are stable
and do not exhibit the GL instability~\cite{Gregory:1994tw}. Hence, the
GL instability also disappears when the Lorentz symmetry along the
world-volume of the black brane is recovered.

Having seen these similarities, it is perhaps tempting to speculate 
that the GL instability of a non-extremal black brane might be
interpreted as the Jeans-like instability of the NG boson associated
with the spontaneous Lorentz breaking along the world-volume of the
black brane. If this is true in some sense then we should expect close
relations between the NG boson $\pi$ in the EFT and the mass density of
the black brane since they are the degrees of freedom exhibiting the
instabilities. Similarly, the spatial components $a_i$ of the $U(1)$
gauge field should be related to the velocity field of the black brane
mass density. To confirm these expectations is far beyond the scope of 
this paper, but it is certainly worthwhile investigating these expected
relations by detailed analysis.

In summary, if the gauge coupling in the gauged ghost condensation is
smaller than a critical value then the NG boson coupled with gravity
exhibits a Jeans-like instability for long wavelength modes as in the
ghost condensation~\cite{Arkani-Hamed:2003uy}. We have pointed out
similarities between the Jeans-like instability of the NG boson and the
Gregory-Laflamme (GL) instability~\cite{Gregory:1993vy,Gregory:1994bj}
of black branes and speculated possible correspondence between these two
instabilities.

\section{Non-extremal black brane and Lorentz breaking}
\label{sec:black-brane}

As already mentioned in the previous section, a black brane can break
the Lorentz symmetry along its world-volume. This is due to different
radial dependence of the time-time and space-space components of the 
metric. Hence, it is tempting to seek a setup in which a black brane
leads to a Higgs phase of gravity in our $4$-dimensional world. For this 
purpose, it is necessary to embed our $4$-dimensional world parallel to
the world-volume of the black brane. This inevitably leads us to
consider braneworld scenarios, in which our $4$-dimensional world is
supposed to be a brane in a higher dimensional bulk spacetime.

In braneworld scenarios, branes other than our world can be included
and some of them might be black branes. Actually, in
(higher-dimensional) general relativity, branes with more than two 
codimensions tend to form black branes in the limit where their
thickness becomes sufficiently small. On the other hand, in the case of 
codimension-$1$, the dynamics of a brane coupled with higher dimensional 
gravity is consistently described by Israel's junction
condition~\cite{Israel:1966rt} even in the thin brane limit. The thin
brane limit of codimension-$2$ objects is more subtle. If the
higher-dimensional geometry surrounding a codimension-$2$ brane is
axisymmetric and if the brane energy momentum is that of vacuum energy,
or tension, then gravity around the brane in the thin brane limit is
consistently described by a deficit angle. However, if either of the two
conditions (the axisymmetry in the bulk and the brane energy momentum of
vacuum energy type) is violated then the description breaks down. For
codimensions more than two, there is no well-defined thin brane limit in
(higher-dimensional) general relativity and a brane tends to form a 
black brane.

As already stated, it often happens that a black brane breaks the
Lorentz symmetry along its world-volume. This is essentially because a
black brane has a preferred frame in which it is at rest and a Lorentz
boost even in the direction parallel to its world-volume does not
preserve the form of the metric. In other words, the time-time component
and the space-space components of the metric depend differently on the
radial coordinate.

In Appendix~\ref{app:four-para-sol} we consider a simple example in
which we can explicity see the Lorentz breaking along the world-volume
of a black brane. We investigate a four-parameter family
of $p$-brane solutions ($p=3,4,5,6$) in type II A/B supergravity. It is
shown that the regularity of the horizon sets two independent
constraints on the parameters of the solutions and, thus, reduces the
four-parameter family of solutions to a two-parameter family. As a
result, a regular non-extremal black $p$-brane in the family of
solutions always breaks the ($p+1$)-dimensional Lorentz symmetry along
its world-volume. On the other hand, the $p$-dimensional spatial
rotational invariance and the ($p+1$)-dimensional translational
invariance are preserved. Intriguingly, the Lorentz symmetry is restored
if and only if the black brane becomes extremal and BPS-saturated.

We expect that this connection between the non-extremality and the
Lorentz breaking should hold in more general situations. Thus, a
non-extremal black brane may be used as a source of spontaneous Lorentz
breaking and may lead to a Higgs phase of gravity.

\section{Setup in string theory}
\label{sec:string}

M/string theory has been considered as a strong candidate for a unified
theory of fundamental physics. Its mathematical consistency and beauty
have been attracting interest of many physicists. On the other hand, one
of its drawbacks is lack of direct experimental or observational
evidence of such a structure at high energies. Having this situation, it
seems rather natural to turn our eyes to cosmology and look for
cosmological implications of M/string theory since the universe is
supposed to have experienced a high energy epoch at its early stage.

In order to realize realistic cosmological scenarios in string theory,
one of the most important issues is the moduli stabilization. In the
context of the type IIB superstring theory, Kachru, Kallosh, Linde and
Trivedi (KKLT)~\cite{Kachru:2003aw} stabilized all moduli by using
various fluxes and non-perturbative corrections to the moduli
potential. Thus, this is a good starting point for cosmology in string
theory.

In the KKLT setup anti-$D3$-branes play an essential role. Inclusion of
anti-$D3$-branes at the bottom of a warped throat uplifts stable
AdS vacua with negative cosmological constant to meta-stable de Sitter
or Minkowski vacua with positive or zero cosmological constant in a
theoretically controllable way. Without anti-$D3$-branes or other
alternative sources such as a $D7$-brane with non-zero flux inside its
world-volume~\cite{Burgess:2003ic,Dudas:2006gr}, we would end up with a 
negative cosmological constant, which is inconsistent with observations.

In this section we consider a situation in which the anti-$D3$-branes at
the bottom of a warped throat may be described by a black brane and
spontaneously break a part of the Lorentz symmetry along its world volume.

Before going into the setup, however, let us list what we have
learned from various considerations in the previous sections. 
\begin{itemize}
 \item[(a)] The structure of the low energy effective field theory (EFT)
	    of gauged ghost condensation is determined solely by the
	    symmetry breaking pattern. Thus, all we have to do is to
	    find a setup realizing the same symmetry breaking pattern. 
 \item[(b)] Similarities between the Jeans-like instability of the
	    Nambu-Goldstone (NG) boson and the Gregory-Laflamme (GL)
	    instability of black branes suggest using a black brane. 
 \item[(c)] A non-extremal black brane tends to spontaneously break the
	    Lorentz symmetry along its world-volume. 
 \item[(d)] In order for a black brane to break the Lorentz symmetry
	    in our world, our world must be parallel to the world-volume
	    of the black brane. This inevitably leads us to consider a
	    braneworld scenario. 
 \item[(e)] In (higher-dimensional) general relativity, branes with
	    codimensions more than two tend to form black branes when
	    the brane thickness is sufficiently small. The string theory 
	    version of this statement is the correspondence
	    principle for stringy objects and black
	    objects~\cite{Horowitz:1996nw}. 
\end{itemize}
These suggest a braneworld scenario with a black brane parallel to our
brane~\footnote{Similar attempts in the context of higher-dimensional
general relativity require violation of the null energy condition or
inclusion of naked singularities other than branes with codimension
one~\cite{Csaki:2000dm,Cline:2001yt}. Our setup includes branes with
higher codimensions as well as non-perturbative effects, which were not 
included in those attempts.}. Consistent braneworld scenarios in string
theory can be constructed in the KKLT setup of warped flux
compactification, which realizes de Sitter vacua in string theory. Thus,
we shall start with the warped flux compactification and seek a
condition under which the correspondence
principle~\cite{Horowitz:1996nw} states that a black brane should
form. In the setup, the world-volume of the black brane shall be
parallel to a brane representing our world.

We consider the KKLT setup of Type IIB compactification with NS-NS and
R-R fluxes. One begins with a warped throat generated by
fluxes~\cite{Klebanov:2000hb} and glues it to a bulk Calabi-Yau $3$-fold
to have a compact extra dimensions~\cite{Giddings:2001yu}. The volume of
the internal space is stabilized by non-perturbative effects such as 
$D$-instantons~\cite{Witten:1996bn}. Since the $4$-dimensional
cosmological constant for this supersymmetric configuration is negative, 
KKLT~\cite{Kachru:2003aw} adds anti-$D3$-branes at the tip of the 
warped throat and explicitly breaks supersymmetry to uplift the AdS
vacua to meta-stable de Sitter vacua. Even if anti-$D3$-branes are
initially placed at some other places in the internal space, they feel
attractive force towards the tip of the throat because of the non-BPS
nature of the configuration~\cite{Kachru:2002gs}. Hence, the
anti-$D3$-branes fall towards the tip and finally settle there because
of the Hubble friction due to expansion of the $4$-dimensional
universe~\cite{Mukohyama:2005cv}.

The geometry deep inside the warped throat is approximated by the
Klebanov-Strassler (KS) solution~\cite{Klebanov:2000hb}. The bottom of
the KS throat is actually not a point but has the topology 
${\bf R}^4\times S^3$, where ${\bf R}^4$ represents the $4$-dimensional
universe and the radius of the $S^3$ is of order $\sqrt{g_sM}\, l_s$, 
where $M$ is an integer representing a quantized flux around the
$S^3$~\cite{Herzog:2001xk}. Thus, if $g_sM\gg 1$ then the curvature of
the geometry is everywhere small and the supergravity approximation is
justified. Throughout this paper we assume this condition. In order for 
the string perturbative expansion to be valid, we require $g_s\lsim 1$
as well. In summary, we assume that
%
\begin{equation}
 g_s M \gg 1, \quad g_s \lsim 1. 
  \label{eqn:consistency}
\end{equation}

Note that this configuration is only meta-stable. The anti-$D3$-branes
can annihilate with the $D3$ charge induced by the background fluxes via
quantum tunneling. Therefore, the configuration lasts for only a finite
duration although the lifetime can be made longer than the age of the
universe.

\subsection{Moduli stabilization}

For this setup to work it is important to make sure that the KKLT type
moduli stabilization is valid. We have to stabilize both complex
structure moduli and K\"ahler moduli in the system. Since all complex
structure moduli are stabilized by fluxes, in this subsection we
consider stabilizing K\"ahler moduli. In the following, for simplicity
we shall consider only one K\"ahler modulus. We make a comment on the
possibility of having many K\"ahler moduli in the end of
subsection~\ref{subsec:correspondence-principle}.

\begin{figure}
 \begin{center}
  \includegraphics[trim = 0 0 0 0 ,scale=0.8, clip]{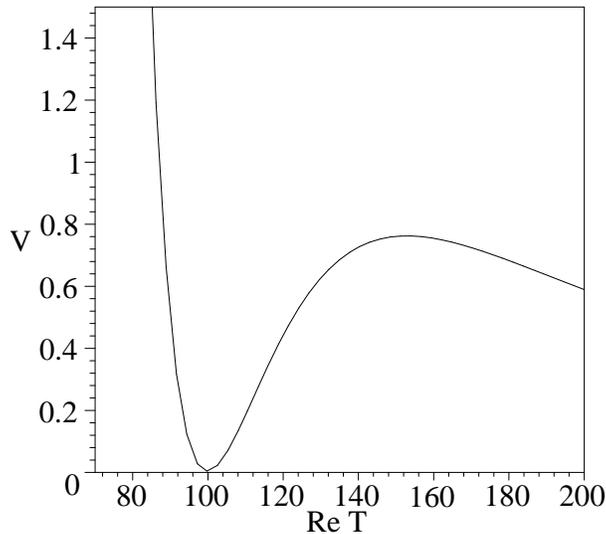}
 \end{center}
 \caption{\label{fig:moduli_potential}
 The moduli potential (multiplied by $10^9$) for $D/A^2=10^{-4}$,
 $W_0/A^2=-0.1107$ and $a=\pi/87$. There is a meta-stable minimum at
 $\sigma\simeq 100$.
 }
\end{figure}

The supersymmetric contribution to the potential for the volume modulus
$\sigma$ combined with the axion $\alpha$ is specified by the K\"ahler
and super potentials of the form 
%
\begin{equation}
 K = -3\log (T+\bar{T}), \quad
  W = W_0 + Ae^{-aT},
\end{equation}
where $T=\sigma+i\alpha$. As in the original KKLT setup, we uplift the
AdS vacua to de Sitter vacua by introducing anti-$D3$-branes, which
adds a non-supersymmetric contribution of the form 
%
\begin{equation}
 \delta V = \frac{D}{(T+\bar{T})^2}, \quad
  D = \frac{2a_0^4T_3\overline{N}_3}{\pi^2g_s^4},
  \label{eqn:deltaV}
\end{equation}
where 
%
\begin{equation}
 a_0 \simeq \exp\left(-\frac{2\pi K}{3g_sM}\right)
  \label{eqn:warp-factor}
\end{equation}
is the warp factor at the bottom of the throat. The total potential for
$\sigma$ (with $\alpha=0$) is 
%
\begin{equation}
 V(\sigma) = 
  \frac{aAe^{-a\sigma}}{2\sigma^2}
  \left(\frac{\sigma aA}{3}e^{-a\sigma}+W_0+Ae^{-a\sigma}\right)
  + \frac{D}{4\sigma^2}. 
\end{equation}
For example, as shown in Fig.~\ref{fig:moduli_potential},
$D/A^2=10^{-4}$ with $W_0/A^2=-0.1107$, $a=\pi/87$ gives a moduli
potential with a meta-stable de Sitter vacuum. For validity of the
geometrical description, it is important to make sure that a local
minimum of the potential is at a sufficiently large value of the volume
modulus $\sigma$. The example shown in Fig.~\ref{fig:moduli_potential}
has a local minimum at $\sigma\simeq 100$, which is large enough.

We would like to make the coefficient $D$ of $\delta V$ sufficiently
small so that the moduli potential with the non-supersymmetric
correction has a local minimum at large enough $\sigma$. On the other
hand, as we shall see in
subsection~\ref{subsec:correspondence-principle}, we would like to 
consider a large $\overline{N}_3$ so that the anti-$D3$-branes 
form a black brane. Since $D\propto a_0^4\overline{N}_3$, the large
$\overline{N}_3$ threatens validity of the moduli stabilization while
the exponential dependence (\ref{eqn:warp-factor}) of $a_0$ on the
fluxes will certainly help reducing $D$. 
%

Note that one cannot make the warp factor $a_0$ arbitrarily small since
the number of branes and fluxes must satisfy the tadpole condition 
%
\begin{equation}
 - \overline{N}_3 + MK  = \frac{\chi}{24},
  \label{eqn:tadpole}
\end{equation}
where $\chi$ is the Euler characteristics of the Calabi-Yau
fourfold~\cite{Giddings:2001yu}. Ref.~\cite{Klemm:1996ts} lists examples
of Calabi-Yau fourfolds in the range 
%
\begin{equation}
 -240 \leq \chi \leq 1820448.\label{eqn:chi-range}
\end{equation}
As far as the author knows, there is no argument prohibiting Calabi-Yau
fourfolds with larger $\chi$. However, for concreteness, we restrict our
considerations to the range (\ref{eqn:chi-range}). The tadpole condition 
(\ref{eqn:tadpole}) combined with the inequalities 
(\ref{eqn:chi-range}) and (\ref{eqn:consistency}) implies that $K$
cannot be arbitrarily large for a fixed $\overline{N}_3$. This means
that the warp factor given by (\ref{eqn:warp-factor}) cannot be
arbitrarily small. Therefore, in order to make $D$ to have a small
value, we must carefully choose parameters ($\overline{N}_3$, $M$, $K$,
$g_s$) so that all consistency conditions are satisfied. In
subsection~\ref{subsec:correspondence-principle} we shall see that this
is indeed possible.

\subsection{A tale of anti-D-branes}
\label{subsec:anti-D-brane}

As already stated, anti-$D3$-branes or other supersymmetry-breaking
branes at the tip of a warped throat is one of the essential ingredients
of the type IIB warped flux compactification. Without those branes, we
would end up with a negative cosmological constant, which is
inconsistent with observations.

The anti-$D3$-branes are distributed over the $S^3$ at the bottom of the
warped throat, but should feel slight attractive force towards each
other since supersymmetry is broken and the attractive gravitational
force is not completely canceled by the repulsive R-R force. Thus, in
the end, they should gather and form a bound object. The gravitational
radius of the bound object made of $\overline{N}_3$ anti-$D3$-branes
should be~\footnote{To be precise, this $R_g$ is the gravitational
radius for a BPS-saturated, isolated stack of anti-$D3$-branes. However,
(\ref{eqn:Rg-anti-D3s}) is expected to be reasonably accurate, provided
that the minimum length scale $\sqrt{g_sM}\, l_s$ of the throat geometry
is sufficiently longer than the $R_g$. This condition is actually
satisfied for the examples (\ref{eqn:parameters-example1}) and
(\ref{eqn:parameters-example2}).}
%
\begin{equation}
 R_g \simeq (4\pi g_s\overline{N}_3)^{1/4}\, l_s. \label{eqn:Rg-anti-D3s}
\end{equation}
It is thought that the system of anti-$D3$-branes should relax to a
non-supersymmetric NS $5$-brane ``giant graviton''
configuration~\cite{Kachru:2002gs}. The NS $5$-brane is wrapped on a 
$S^2$ inside the internal $S^3$ and carries anti-$D3$ charge
$\overline{N}_3$ coming from a world-volume magnetic flux. This
configuration is classically stable for 
%
\begin{equation}
 \frac{\overline{N}_3}{M} < 0.08, \label{eqn:NS5-stability}
\end{equation}
but can quantum-mechanically tunnel to a supersymmetric vacuum. The
decay rate is 
%
\begin{equation}
 \Gamma \sim l_s^{-1}\exp 
  \left(-\frac{27\pi^4b_0^{12}g_sM^6}{64\overline{N}_3^3}\right),
\end{equation}
where $b_0\simeq 0.93266$. Being conservative~\footnote{
If $l_s$ is longer then this condition becomes weaker.}, if we suppose  
$l_s\simeq M_{Pl}^{-1}$  then $\Gamma\ll H_0\simeq 10^{-61}M_{Pl}$
requires that 
%
\begin{equation}
 \frac{g_sM^6}{\overline{N}_3^3} \gsim 8. \label{eqn:NS5-long-lifetime}
\end{equation}
This condition is easily satisfied. For this reason, we consider the NS
$5$-brane state as a starting point of our discussion. The radius of the
$S^2$ on which the NS $5$-brane is wrapped is 
%
\begin{equation}
 R_{S^2} = \frac{2\pi\overline{N}_3}{b_0^3M}\, \sqrt{g_sM}\, l_s.
  \label{eqn:R-S2}
\end{equation}

\subsection{What does the correspondence principle require?}
\label{subsec:correspondence-principle}

We shall now seek the condition under which the anti-$D3$-branes at the
tip of the warped throat should be described by a non-extremal 
black brane, based on the correspondence principle for $D$-branes and
black branes~\cite{Horowitz:1996nw}.

Roughly speaking, the correspondence principle says that a stringy
object and the corresponding black object with the same charges are
different descriptions of the same object. The black object is a better
description if the gravitational radius is larger than the size of the
stringy object. Thus, in order for the anti-$D3$-branes to be described
by a black brane, the gravitational radius $R_g$ given by
(\ref{eqn:Rg-anti-D3s}) must be greater than both the string scale $l_s$
and the radius $R_{S^2}$ of the non-supersymmetric NS $5$-brane ``giant 
graviton'' configuration given by (\ref{eqn:R-S2}). Therefore,
hereafter, we impose the condition 
%
\begin{equation}
 R_g \gsim R_{S^2}, \quad R_g \gsim l_s.
  \label{eqn:gsMgsNbar}
\end{equation}
Under this condition, according to the correspondence principle for
$D$-branes and black branes~\cite{Horowitz:1996nw}, the bound object of
$\overline{N}_3$ anti-$D3$-branes should be described by a black
brane. Note, however, that this configuration cannot be a BPS saturated
state since supersymmetry is explicitly broken in the present
setup. Therefore, we conclude that this must be a non-extremal black
brane.

We now see that there is a range of parameters in which the condition
(\ref{eqn:gsMgsNbar}) is compatible with the KKLT-type moduli
stabilization. Again, for concreteness, we adopt the range
(\ref{eqn:chi-range}) for the Euler characteristics of the Calabi-Yau
fourfold. The goal is to find a set of parameters ($\overline{N}_3$,
$M$, $K$, $g_s$) which satisfies (\ref{eqn:gsMgsNbar}),
(\ref{eqn:NS5-stability}), (\ref{eqn:NS5-long-lifetime}),
(\ref{eqn:consistency}) and (\ref{eqn:chi-range}) simultaneously and
which gives an sufficiently small value of the coefficient $D$ of
$\delta V$. Here, $\chi$ should satisfy the tadpole condition
(\ref{eqn:tadpole}) and $D$ is given by the formula
(\ref{eqn:deltaV}). This is indeed possible. For example, 
%
\begin{equation}
 \overline{N}_3 = 22, \quad
  M = 553, \quad
  K = 131, \quad
  g_s = 0.1 \label{eqn:parameters-example1}
\end{equation}
satisfy all conditions, leading to 
%
\begin{equation}
 a_0 \simeq 7\times 10^{-3}, \quad
 \frac{D}{T_3} \simeq 10^{-4}, \quad
  \chi = 1738104,
\end{equation}
and
%
\begin{equation}
 \frac{R_{S^2}^2}{R_g^2} \simeq 1, \quad
  \frac{l_s^2}{R_g^2} \simeq 0.19\ . 
\end{equation}
This value of $D$ is small enough that the anti-$D3$-brane contribution
to the moduli potential is under control. Indeed, this value roughly
corresponds to the value used in Fig.~\ref{fig:moduli_potential} if
$A^2\simeq T_3$. Another example is 
%
\begin{equation}
 \overline{N}_3 = 14, \quad
  M = 397, \quad
  K = 180, \quad
  g_s = 0.2, \label{eqn:parameters-example2}
\end{equation}
which corresponds to 
%
\begin{equation}
 a_0 \simeq 9\times 10^{-3}, \quad
 \frac{D}{T_3} \simeq 10^{-5}, \quad
  \chi = 1714704,
\end{equation}
and
%
\begin{equation}
 \frac{R_{S^2}^2}{R_g^2} \simeq 1, \quad
  \frac{l_s^2}{R_g^2} \simeq 0.17\ . 
\end{equation}
Note that there is no argument prohibiting $\chi$ larger than
(\ref{eqn:chi-range}). If Calabi-Yau fourfolds with larger $\chi$ are
found then it will be easier to satisfy all the consistency conditions.

In summary, if the number of anti-$D3$-branes $\overline{N}_3$, the flux 
number $M$ and the string coupling $g_s$ satisfy the condition
(\ref{eqn:gsMgsNbar}) then the type IIB warped flux compactification
should include a non-extremal black brane at the bottom of a warped
throat. As shown explicitly, it is possible to find a set of parameters
for which the condition (\ref{eqn:gsMgsNbar}) is compatible with the
KKLT-type moduli stabilization.

Note that the gravitational radius $R_g$ in the above examples is not
large enough to make classical description of the black brane
reliable. Indeed, since $l_s^2/R_g^2 \sim 20\%$, we should expect
$\alpha'$-corrections of order $20\%$ or so. We suppose that those 
$\alpha'$-corrections do not accidentally restore the symmetry broken in 
the classical level and that the Lorentz symmetry along the world-volume
of the black brane remains spontaneously broken. In the next subsection
we shall introduce another brane parallel to the world-volume of the
black brane and consider it as our $4$-dimensional universe. At the
position of our brane the $\alpha'$-corrections should be negligible and
the geometry can be treated classically if it is sufficiently far from
the black brane.

Note also that, while we have considered stabilizing only one K\"ahler
modulus, in general the number of K\"ahler moduli is not just one. 
Indeed, Calabi-Yau fourfolds with large Euler number, which we have
considered, tend to have many K\"ahler moduli. As a result, rigorous
treatment of K\"ahler moduli stabilization is rather involved. (See e.g.
\cite{Denef:2005mm}, where $51$ K\"ahler moduli are stabilized.) In the
above analysis, following KKLT and other works in the literature, we
have considered only one K\"ahler modulus and have seen how difficult it
is to achieve stabilized models. In particular, we have seen that the
moduli stabilization sets strong constraints on the model parameters
($\overline{N}_3$, $M$, $K$, $g_s$) other than those directly related to
the K\"ahler moduli stabilization ($W_0$, $A$, $a$). Since inclusion of
all other K\"ahler moduli certainly complicates the analysis, at this
moment we do not know whether inclusion of them strengthens or weakens
the constraints on the former set of model parameters ($\overline{N}_3$,
$M$, $K$, $g_s$). With a large number of K\"ahler moduli, we will
certainly have extra parameters as well as extra stability
conditions. Further studies towards complete analysis of moduli
stabilization are certainly worthwhile pursuing.

\subsection{Lorentz breaking on our brane}

We now introduce another brane as our universe and arrange it to be
parallel to the world-volume of the black brane. Our brane may be placed
either in the bulk region, where the warp factor is of order unity, or
in another throat.

If the KKLT type moduli stabilization is valid then this configuration
allows the induced geometry on our brane to be maximally symmetric,
either Minkowski or de Sitter. Judging from what happens in the
Randall-Sundrum type braneworld
cosmology~\cite{Maartens:2000fg,Csaki:1999jh,Cline:1999ts,Flanagan:1999cu,Binetruy:1999hy,Mukohyama:1999qx,Kraus:1999it,Ida:1999ui,Mukohyama:1999wi},  
one might expect that the black brane would behave as ``dark
radiation''~\cite{Mukohyama:1999qx} and that the $4$-dimensional 
universe would expand like radiation-dominated one. However, since
all moduli are stabilized, this expectation is not correct and the black
brane is indistinguishable from the $4$-dimensional cosmological
constant from the viewpoint of the gravitational source driving the 
homogeneous, isotropic evolution of the $4$-dimensional universe. In the 
Randall-Sundrum type setup without radion stabilization, the scale
factor of the $4$-dimensional universe on the UV brane would be the
radial position of the UV brane in the $5$-dimensional AdS-Schwarzschild 
geometry~\cite{Mukohyama:1999wi}. In this language the expansion of the
universe would be due to the motion of the brane away from the black 
brane. The effect of the black brane on the UV brane would be diluted as
the UV brane moves away, and this is the reason why the effective energy 
density of the "dark radiation" scales as $1/a^4$ in accord with the
radial dependence of the projected Weyl tensor~\cite{Shiromizu:1999wj}. 
On the other hand, in the present setup all moduli are stabilized. Thus,
the shape and the size of the internal space do not change significantly
as the universe expands. Since both the black brane and our brane are
located in the internal space, this means that there is no way to dilute
the influence of the black brane on our $4$-dimensional universe. The
black brane is always there! (The Hawking radiation is negligible for
tiny deviation from extremality.) Thus, we are led to the conclusion
that the effective $4$-dimensional energy density induced by the black
brane cannot depend as strongly as the dark radiation on the scale
factor of the $4$-dimensional universe. In other words, the effect of
the black brane on our brane is indistinguishable from a cosmological
constant as far as homogeneous, isotropic evolution of the
$4$-dimensional universe is concerned. Therefore, the $4$-dimensional
universe should admit maximally symmetric solutions, either Minkowski or
de Sitter spacetime.

The non-extremal black brane breaks the Lorentz symmetry along the
world-volume down to the spatial rotational invariance and the spacetime
translational invariance. This is due to different radial dependence of
the time-time and the space-space components of the black brane
metric. Intuitively, this is because the black brane has a preferred
frame in which it is at rest. Since the brane representing our
$4$-dimensional world is parallel to the world-volume of the black
brane, the Lorentz symmetry in our $4$-dimensional world is
spontaneously broken by the existence of the black brane but the spatial
rotational invariance and the spacetime translational invariance are
still preserved. Note again that the induced metric on our brane is
Minkowski or de Sitter even though the Lorentz symmetry is broken. This
leads to a Higgs phase of gravity in our world.

\subsection{Horizon modes and (locally) localized gravity}

Suppose that the warp factor on our brane is much larger than the small
warp factor near the black brane. This is indeed the case if our brane
is located in the bulk region, where the warp factor is of order unity,
or in another throat shorter than the black brane throat.

In this case, due to the (locally) localized gravity
phenomenon~\cite{Randall:1999vf,Karch:2000ct}, the overlap of the
graviton zero mode with modes localized near the black brane horizon is
exponentially suppressed compared with the overlap with matter on our
brane. This is because the overlap is controlled by the warp
factor. Therefore, gravity on our brane is essentially unaffected by
modes localized near the black brane.

The black brane horizon should change the boundary condition for
$10$-dimensional fields near the tip of the warped throat and modify the
Kaluza-Klein (KK) spectrum significantly. In a sense, this is analogous
to what happens in the 't Hooft's brick wall model for black hole
entropy~\cite{'tHooft:1984re,Mukohyama:1998rf}. Because of the infinite
blue-shift at the horizon, modes with finite frequencies measured from a
distance can have arbitrarily high local frequencies near the
horizon. Therefore, there can be infinite number of extra modes
associated with the presence of the horizon. The 't Hooft's proposal was
to relate the number of those extra modes to the black hole entropy. In
the present setup, one might think that those extra modes threaten the
recovery of the $4$-dimensional gravity since they are expected to form
a gapless mass spectrum.

Nonetheless, those extra light modes localized near the black brane do
not affect gravity on our brane. We are considering a situation where
the warp factor on our brane is much larger than the warp factor near
the black brane. Therefore, as explained above, due to the (locally)
localized gravity, those extra modes essentially do not interact with
the graviton zero mode nor with matter on our brane. Thus,
$4$-dimensional gravity should be recovered on our brane.

The same reasoning implies that the scale of spontaneous Lorentz
breaking on our brane should be rather low compared with the fundamental
scale. This is because Lorentz breaking modes are localized near the
black brane and, thus, has exponentially small overlap with the graviton
zero mode. Despite the tininess of Lorentz breaking, the Higgs phase of
gravity has dramatic consequences on gravity and cosmology in our world 
as reviewed in the introduction.

\section{Summary and discussion}
\label{sec:summary}

As a step towards a Higgs phase of gravity in string theory, we have
considered a braneworld scenario with a black brane parallel to our
brane. This is motivated by the following observations. 
(a) The structure of the low energy effective field theory (EFT) of
gauged ghost condensation is determined solely by the symmetry breaking
pattern. 
(b) Similarities between the Jeans-like instability of the
Nambu-Goldstone (NG) boson and the Gregory-Laflamme (GL) instability of
black branes suggest using a black brane. 
(c) A non-extremal black brane tends to spontaneously break the Lorentz
symmetry along its world-volume. 
(d) In order for a black brane to break the Lorentz symmetry in our
world, our world must be parallel to the world-volume of the black
brane. 
(e) In (higher-dimensional) general relativity, branes with codimensions
more than two tend to form black branes when the brane thickness is
sufficiently small.

The existence of the black brane horizon spontaneously breaks the
Lorentz symmetry on our brane but preserves the rotational and
translational invariance. To investigate moduli stabilization, we have
considered a KKLT-type moduli potential and found sets of parameters
which realize stabilized moduli. We have also pointed out a possible
obstacle to the use of the KKLT-type moduli potential which stabilizes
only one K\"ahler modulus: our setup tends to include many K\"ahler
moduli and the complete analysis should be rather involved. If the
moduli stabilization remains valid after taking all K\"ahler moduli into
account, then this setup leads to a Higgs phase of gravity in string
theory and, thus, may be considered as a UV completion of the gauged
ghost condensation.

If the gauge coupling in the EFT of the gauged ghost condensation is
small enough then this setup reduces to the ghost condensation and the
NG boson coupled to gravity exhibits Jeans-like instability. We have
speculated that the geometrical counter-part of Jeans-like instability
might be related to the GL instability of the non-extremal black brane.

Having proposed a possible scenario towards a UV completion of the
gauged ghost condensation, let us now discuss physics beyond the low
energy EFT. For example, this geometrical setup may make it possible for
us to think about transition from the symmetric phase to the broken
phase. As described in  subsection~\ref{subsec:anti-D-brane},
$\overline{N}_3$ anti-$D3$-branes  distributed over the $S^3$ at the
bottom of the warped throat feel attractive force towards each other and
they form a bound object. Provided that $g_s\lsim 1$ and that
$g_s\overline{N}_3\gsim 1$, initially the gravitational radius of each
anti-$D3$-brane is less than the string scale but the gravitational
radius of the final bound object is greater than the string
scale. Therefore, as argued in
Sec.~\ref{subsec:correspondence-principle}, while the stringy picture is
a good description for the initial state, the black brane picture is 
more relevant for the final state. Thus, this process can be considered
as gravitational collapse of a collection of anti-$D3$-branes to form a
black brane. Before the gravitational collapse, the $4$-dimensional
Lorentz symmetry is preserved. On the other hand, the $4$-dimensional
Lorentz symmetry is spontaneously broken after the formation of the
black brane by gravitational collapse.

In this picture, the final state is stationary. This is the reason why
the residual symmetry can include not only the $3$-dimensional spatial 
diffeomorphism invariance but also an unbroken $U(1)$ gauge symmetry. In
the limit $g^2\ll g_c^2$, the latter symmetry is translated to a global
shift symmetry in the EFT of the NG boson. Therefore, it is easy to
infer how we can break the shift symmetry. In the situation where
$g^2\ll g_c^2$, if the non-extremal black brane at the tip of the warped
throat is not exactly stationary but time-dependent then the shift
symmetry is broken, while the $3$-dimensional spatial diffeomorphism
invariance still remains. If the black brane is quasi-stationary then
the breaking of the shift symmetry should be very weak.

It is certainly worthwhile considering other nonlinear and/or UV issues
beyond the EFT. As an example, the UV completion might provide some
new insight on nonlinear dynamics of the NG boson triggered by the
Jeans-like instability for the case $g^2<g_c^2$. The endpoint of the GL
instability is still a question under debate. It might be a small black
holes or a non-uniform black brane. In any case, according to our
consideration, the endpoint of the GL instability should correspond to
the endpoint of nonlinear evolution of the NG boson triggered by the
Jeans-like instability. Therefore, knowledge about the endpoint of the
GL instability might provide new insight on nonlinear dynamics of
gravity in the Higgs phase. In particular, this line of consideration
might determine the non-linear properties of an alternative to dark
matter in the context of ghost condensation.

The way to end the ghost inflation is also an issue for which the UV
completion might be useful. In ref.~\cite{Arkani-Hamed:2003uz} it was 
assumed that the shift symmetry is broken locally so that an inflationary
de Sitter phase ends gracefully. In the EFT language, this can, for
example, be due to a phase transition in another sector triggered by 
the scalar field responsible for the ghost condensate. In our UV
completion, as discussed above, the shift symmetry is broken if the black
brane at the bottom of a warped throat is time-dependent. Thus, if the
black brane experiences a transition from an almost stationary phase to
another almost stationary phase then the shift symmetry is broken
essentially in the transition period only. Depending on the nature of
the transition, the $4$-dimensional effective cosmological constant
changes and the initial inflationary phase can end gracefully. The cause
of the transition may be a capture of another brane by the black brane,
a merger of black branes, and so on. There may be other possibilities to
UV complete the idea of ghost inflation.

Couplings to the matter sector is also worthwhile investigating in
the geometrical setup. In the EFT language,
refs.~\cite{Arkani-Hamed:2004ar,Cheng:2006us} listed allowed
couplings between the (ungauged or gauged) ghost condensate and the 
matter sector, and discussed physical consequences. For example, the
Lorentz- and CPT-violating Chern-Simon operator of the electromagnetic
field can be forbidden by imposing a discrete
symmetry~\cite{Arkani-Hamed:2004ar}. It is certainly important to see if
the matter sector can be embedded with this kind of discrete symmetry in
the present geometrical setup.

As already stated many times, the ghost condensation can be obtained as
the $g^2\ll g_c^2$ limit of the gauged ghost condensation, for which we
have suggested a possible UV completion. In the context of the EFT, this
limit can be achieved by fine-tuning the gauge coupling $g$. However, we
do not yet know whether this limit of the geometrical setup really
exists or not. It is certainly worthwhile to express actual values of
the coupling constant $g$ and the scale of symmetry breaking ${\cal M}$
in terms of the parameters of the geometrical setup.

\section*{Acknowledgements}

The author is grateful to M.~A.~Luty for helpful discussions and
continuing encouragement. He would like to thank N.~Arkani-Hamed, 
H.-C.~Cheng, S.~Dubovsky, G.~Dvali, Y.~Imamura, E.~Kiritsis, A.~Nicolis,
N.~Ohta, R.~Sriharsha and T.~Wiseman for useful comments. He
acknowledges warm hospitality at Maryland Center for Fundamental Physics
and Harvard High Energy Theory Group, where a part of this work was
done. This work was in part supported by MEXT through a Grant-in-Aid for
Young Scientists (B) No.~17740134.

\appendix{Appendices}

\subsection{Four-parameter family of black-brane solutions}
\label{app:four-para-sol}

We consider the Einstein-frame metric $ds_{10}^2$, the dilaton $\phi$
and the RR field $C^{(p+1)}$ of the form
%
\begin{eqnarray}
 ds_{10}^2 & = & -e^{2A_0(r)}dt^2 + e^{2A(r)}\delta_{ij}dx^idx^j 
  + e^{2B(r)}(dr^2+r^2\Omega^{(8-p)}_{mn}d\theta^md\theta^n), \nonumber\\
 \phi & = & \phi(r), \nonumber\\
 C^{(p+1)} & = & -C(r)\  dt\prod_{i=1}^{p}\wedge dx^i, 
\end{eqnarray}
where $i,j=1,\cdots,p$; $m,n=p+2,\cdots,9$; and
$\Omega^{(8-p)}_{mn}d\theta^md\theta^n$ is the metric of the unit
($8-p$)-sphere. The RR field $C^{(p+1)}$ is related to the RR field
strength $F_{8-p}$ as $F_{8-p}=e^{(3-p)\phi/2}*dC^{(p+1)}$ for
$p=4,5,6$ and to the self-dual RR field strength $F_5$ as
$F_5=(1/\sqrt{2})(dC^{(4)}+*dC^{(4)})$ for $p=3$. In order to impose the
regularity of the black brane horizon, the following expression for the
Riemann tensor is useful. 
%
\begin{eqnarray}
 R^{tr}_{\ \ tr} & = & -(A_0''+{A_0'}^2-A_0'B')e^{-2B}, \nonumber\\
 R^{ir}_{\ \ jr} & = & -(A''+{A'}^2-A'B')e^{-2B}\delta^i_j, \nonumber\\
 R^{mr}_{\ \ \ nr} & = & -\left(B''+\frac{B'}{r}\right)e^{-2B}\delta^m_n, 
  \nonumber\\
 R^{mn}_{\ \ \ m'n'} & = & 
  -B'\left(B'+\frac{2}{r}\right)e^{-2B}
      (\delta^m_{m'}\delta^n_{n'}-\delta^m_{n'}\delta^n_{m'}), 
  \nonumber\\
 R^{ti}_{\ \ tj} & = & -A_0'A'e^{-2B}\delta^i_j, \nonumber\\
 R^{tm}_{\ \ tn} & = &
  -A_0'\left(B'+\frac{1}{r}\right)e^{-2B}\delta^m_n, 
  \nonumber\\
 R^{im}_{\ \ \ jn} & = &
  -A'\left(B'+\frac{1}{r}\right)e^{-2B}\delta^i_j\delta^m_n, 
  \nonumber\\
 R^{ij}_{\ \ i'j'} & = &
  -{A'}^2e^{-2B}(\delta^i_{i'}\delta^j_{j'}-\delta^i_{j'}\delta^j_{i'}). 
\end{eqnarray}

The general solution to the type II A/B supergravity with this ansatz is
obtained as a four-parameter family of $p$-brane
solutions~\cite{Zhou:1999nm}. In the notation of \cite{Brax:2000cf} the 
four-parameter family is written as
%
\begin{eqnarray}
A_0(r) & = & A(r) + \frac{1}{2}\ln f(r), \nonumber\\
f(r) &=& e^{- c_3 h(r)}, \nonumber\\
 A(r) & = &  \frac{(7-p)}{32}\left\{ -\frac{p-3}{2}c_1
+ \left[1 + \frac{(p-3)^2}{8(7-p)}\right]\, c_3 \right\}\, h(r) 
\nonumber \\
& & 
 - \frac{7-p}{16}\ln
 \left[\cosh(k \, h(r) ) - c_2\, \sinh( k \, h(r) ) \right],\nonumber\\
B(r) & = &    \frac{1}{7-p} \, \ln \left[f_-(r) f_+(r) \right]
 +\frac{p-3}{64} \left[(p+1)c_1 +\frac{p-3}{4}c_3 \right]\, h(r) 
\nonumber \\
& &  + \frac{p+1}{16}\, 
 \ln\left[ \cosh( k \, h(r) ) - c_2\, \sinh( k \, h(r) ) \right], 
 \nonumber\\
\phi(r) & = &    \frac{7-p}{16}
\left[(p+1)c_1 +\frac{p-3}{4}c_3 \right]\, h(r)   
\nonumber\\
& & -\frac{p-3}{4} \, 
 \ln\left[ \cosh(k \, h(r) ) - c_2\, \sinh( k \, h(r) ) \right],
 \nonumber\\
 C(r) & = & \pm \, \frac{\sqrt{c_2^2-1}\sinh (k \, h(r) )} 
  {\cosh( k \,  h(r) )  - c_2\, \sinh( k \, h(r) ) }, 
\end{eqnarray}
where
%
\begin{eqnarray}
h(r)  &=& \ln \left[\frac{f_-(r)}{f_+(r)} \right] \ , 
\nonumber\\
f_\pm(r) &\equiv & 1\pm \left(\frac{r_0}{r}\right)^{7-p} \ ,
\nonumber\\
k & = & \sqrt{{2(8-p) \over 7-p }
- c_1^2 + \frac{1}{4}\left(\frac{3-p}{2}\ c_1 +
\frac{7-p}{8}\ c_3 \right)^2 - \frac{7}{16}\ c_3^2}.
\end{eqnarray}
The reality and regularity of the RR field in the region
$|r_0^{7-p}|\leq r^{7-p}<\infty$ require that $k$ and $c_2$ be real and 
that $c_2\geq 1$ and $c_2\leq -1$ for $r_0^{7-d}> 0$ and $r_0^{7-d}< 0$,
respectively. In order for $k$ to be real, $c_1$ and $c_3$ must satisfy
$c_{1-}\leq c_1\leq c_{1+}$ and $c_3^2\leq c_{3max}^{\ \ 2}$, where 
%
\begin{eqnarray}
 c_{1\pm} & = & -\, \frac{(p-3)c_3}{4(p+1)}\pm \frac{2}{p+1}
  \sqrt{\frac{2p(c_{3max}^{\ \ 2}-c_3^2)}{7-p}}, \nonumber\\
 c_{3max}^{\ \ 2} & = & \frac{4(p+1)(8-p)}{p(7-p)}. 
\end{eqnarray}

Let us first consider the case with $r_0^{7-d}> 0$ (we shall consider
the case with $r_0^{7-p}<0$ later). In this case, for generic values of
the parameters the Ricci scalar diverges at $r=r_0$ ($>0$). The leading
term is $R\sim \beta_Rr_0^{-2}y^{-\gamma}$, where 
%
\begin{eqnarray}
 y & = & \frac{7-p}{2}\left(\frac{r}{r_0}-1\right), \nonumber\\
 \gamma & = & 
 -\left[
   \frac{(p+1)(p-3)c_1}{32}+\frac{(p-3)^2c_3}{128}+\frac{2(8-p)}{7-p}\right]
 + \frac{p+1}{8}k, \nonumber\\
 \beta_R & = & 2^{-[10+4/(7-p)-(p+1)/8]}(c_2+1)^{-(p+1)/8}
  (7-p)^3\times\nonumber\\
 & & 
  \left\{\frac{16(8-p)(p-3)^2}{(7-p)^2}
   - (p+1)(p^2-6p+1)c_1^2-\frac{(p-3)(p^2-6p+1)}{2}c_1c_3 
  \right.\nonumber\\
 & & \left.
      +\frac{(p-3)^2(p^2-14p-7)}{16(7-p)}c_3^2      
      +(p-3)[4(p+1)c_1+(p-3)c_3]k\right\}.
 \label{eqn:y-betaR-gamma}
\end{eqnarray}
It is easy to show that $\gamma$ is positive and, thus, $R$ indeed
diverges at $r=r_0$ unless $\beta_R=0$. By setting $\beta_R=0$, we
obtain 
%
\begin{equation}
 c_1 = -\frac{(p-3)c_3}{4(p+1)}
  - \frac{p-3}{p+1}\sqrt{\frac{p(c_{3max}^{\ \ 2}-c_3^2)}{2(7-p)}}.
  \label{eqn:c1-value}
\end{equation}
With this value of $c_1$, the leading behavior of the Kretchmann scalar 
$K\equiv R^{MN}_{\ \ \ M'N'}R^{M'N'}_{\ \ \ \ MN}$ near $r=r_0$ is 
$K\sim \beta_Kr_0^{-4}y^{-2\gamma}$, where 
%
\begin{eqnarray}
 \beta_K & = &  2^{-[10+8/(7-p)-(p+1)/4]}(c_2+1)^{-(p+1)/4}\, 
  \frac{p(7-p)^3}{p+1}\nonumber\\
 & & 
  \times\left\{\frac{7p^3-74p^2+375p-56}{(p+1)^2}c_3^4-\frac{64(8-p)(p-1)}{p+1}c_3^3
   \right.\nonumber\\
 & & 
  -\frac{8(p+17)(8-p)}{p+1}c_3^2
   -\frac{16(8-p)(23p^3-386p^2+1319p+2304)}{p(7-p)^2}\nonumber\\
 & & \left.+128
      \left[ \frac{4(p-1)}{(p+1)^2}c_3^3+\frac{(15-p)(8-p)}{(7-p)(p+1)}c_3^2
       + \frac{4(p+9)(8-p)^2}{p(7-p)^2}\right]k
      \right\}, \nonumber\\
 k & = &
  \frac{1}{8}\sqrt{2p(7-p)\left(c_{3max}^{\ \ 2}-c_3^2\right)}. 
  \label{eqn:betaK}
\end{eqnarray}
By requiring that $\beta_K=0$, we obtain~\footnote{
For $p=1$, $c_3=2$ also satisfies $\beta_K=0$. However, 
we need $p\geq 3$ to accommodate the $3$-dimensional spatial
diffeomorphism of our $4$-dimensional universe. Thus, we restrict our
consideration to the cases with $p=3,4,5,6$. In these cases $C_2=-2$ is
the unique solution to $\beta_K=0$.} $c_3=-2$. Therefore, the regularity
of the horizon at $r=r_0$ requires that 
%
\begin{equation}
 c_1= -\, \frac{p-3}{2(7-p)}, \quad c_3 = -2 \quad
  \mbox{for } r_0^{7-d}>0. 
\end{equation}

The above analysis can be repeated for the case with $r_0^{7-p}<0$. The
leading behavior of the Ricci scalar near $r=|r_0|$ is 
$R\sim\tilde{\beta}_R|r_0|^{-2}y^{-\tilde{\gamma}}$, where
$\tilde{\beta}_R$ and $\tilde{\gamma}$ are $\beta_R$ and $\gamma$ in
(\ref{eqn:y-betaR-gamma}) with the replacement
$(c_1,c_2,c_3)\to(-c_1,-c_2,-c_3)$. Again, $\tilde{\gamma}$ is shown to 
be positive and the regularity of the Ricci scalar at $r=|r_0|$ requires
$\tilde{\beta}_R=0$, which implies (\ref{eqn:c1-value}) with the
replacement $(c_1,c_3)\to(-c_1,-c_3)$. With this value of $c_1$, the
leading behavior of the Kretchmann scalar is 
$K\sim\tilde{\beta}_K|r_0|^{-4}y^{-2\tilde{\gamma}}$, where
$\tilde{\beta}_K$ is $\beta_K$ in (\ref{eqn:betaK}) with the replacement 
$(c_2,c_3)\to(-c_2,-c_3)$. Hence, by requiring $\tilde{\beta}_K=0$, we
obtain
%
\begin{equation}
 c_1= \frac{p-3}{2(7-p)}, \quad c_3 = 2 \quad
  \mbox{for } r_0^{7-d}<0. 
\end{equation}

We are now left with the two parameters $r_0^{7-d}$ and $c_2$. The 
$10$-dimensional Einstein-frame metric is 
%
\begin{equation}
 ds_{10}^2 = e^{2A(r)}
  \left[- \left(\frac{\bar{f}_-(r)}{\bar{f}_+(r)}\right)^2dt^2 
   + \delta_{ij}dx^idx^j \right]
  + e^{2B(r)}(dr^2+r^2\Omega^{(8-p)}_{mn}d\theta^md\theta^n),
\end{equation}
where
%
\begin{eqnarray}
 \bar{f}_{\pm}(r) & = & 1\pm \left(\frac{|r_0|}{r}\right)^{7-p}, \nonumber\\
 A(r) & = & -\frac{7-p}{16}
  \ln
  \left[\frac{1-|c_2|}{2}\left(\frac{\bar{f}_-(r)}{\bar{f}_+(r)}\right)^2+\frac{1+|c_2|}{2}\right],
  \nonumber\\
 B(r) & = & \frac{1}{7-p}\ln\left(\bar{f}_-(r)\bar{f}_+(r)\right)
  - \frac{(p-3)^2}{16(7-p)}\ln \left(\frac{\bar{f}_-(r)}{\bar{f}_+(r)}\right) \nonumber\\
 & & + \frac{p+1}{16}\ln
  \left[\frac{1-|c_2|}{2}\left(\frac{\bar{f}_-(r)}{\bar{f}_+(r)}\right) 
   +\frac{1+|c_2|}{2}\left(\frac{\bar{f}_+(r)}{\bar{f}_-(r)}\right)\right].
\end{eqnarray}
The dilaton and the RR field are 
%
\begin{eqnarray}
 \phi(r) & = & -\frac{p-3}{4}
  \ln
  \left[\frac{1-|c_2|}{2}\left(\frac{\bar{f}_-(r)}{\bar{f}_+(r)}\right)^2+\frac{1+|c_2|}{2}\right],
  \nonumber\\
 C(r) & = & \pm 
  \frac{\sqrt{|c_2|^2-1}\left[\bar{f}_+(r)^2-\bar{f}_-(r)^2\right]}
  {(1+|c_2|)\bar{f}_+(r)^2+(1-|c_2|)\bar{f}_-(r)^2}.
\end{eqnarray}
Note that $c_2\geq 1$ for $r_0^{7-p}>0$ and $c_2\leq -1$ for
$r_0^{7-p}<0$. The two parameter solution actually agrees with the black 
$p$-brane solution of ref.~\cite{Horowitz:1991cd} up to the coordinate
transformation between the isotropic coordinate in the present analysis
and the Schwarzschild-like coordinate in \cite{Horowitz:1991cd}. From
this form of the metric, it is seen that, for $r_0^{7-p}\ne 0$, the
solution has a $p$-dimensional spatial rotational invariance but does
not have a ($p+1$)-dimensional Lorentz symmetry. For $r_0^{7-p}=0$ with
$c_2$ finite, the solution is trivial.

If we take the scaling limit $r_0^{7-p}\to \pm 0$, $c_2\to\pm\infty$
with $c_2\, r_0^{7-p}$ kept finite, the solution reduces to the BPS
$Dp$-brane: 
%
\begin{eqnarray}
 ds_{10}^2 & = & f_0^{-\frac{7-p}{8}}\, \eta_{\mu\nu}dx^{\mu}dx^{\nu}
  + f_0^{\frac{p+1}{8}}(dr^2+r^2\Omega^{(8-p)}_{mn}d\theta^md\theta^n), \nonumber\\
 e^{\phi} & = & f_0^{-\frac{p-3}{4}}, \nonumber\\
 C & = & \pm \left( 1 - f_0^{-1}\right),
\end{eqnarray}
where $\mu,\nu=0,1,\cdots,p$, $x^0=t$ and 
%
\begin{equation}
 f_0 = 1 + \frac{\mu_0}{r^{7-p}}, \quad \mu_0 \equiv 2 c_2r_0^{7-p}.
\end{equation}
In this limit the ($p+1$)-dimensional Lorentz symmetry is restored but
the extremality is also recovered at the same time.

On the other hand, for generic values of the parameters $r_0^{7-d}$ and
$c_2$, the solution does not possess the ($p+1$)-dimensional Lorentz
symmetry. Therefore, we conclude that within the four-parameter family
of solutions, a regular non-extremal black $p$-brane always breaks the
($p+1$)-dimensional Lorentz symmetry along its world-volume but
preserves the $p$-dimensional spatial rotational invariance. Evidently,
the ($p+1$)-dimensional translational invariance along the world-volume
is unaffected by the non-extremality.


\end{document}